\documentclass[showpacs,amsmath,amssymb,aps,pra,preprint,groupedaddress]{revtex4-1}

\usepackage[latin1]{inputenc}

\begin{document}


\title{A list of identities made with products between two different generators of the Clifford algebra}


\author{J. B. Formiga}
\email[]{jansen.formiga@uespi.br}
\affiliation{Centro de Ciências da Natureza, Universidade Estadual do Piauí, C. Postal 381, 64002-150 Teresina, Piauí, Brazil}


\date{\today}

\begin{abstract}
Here I present a full list with all possibles products between the generators of the Clifford algebra in a four-dimensional spacetime. The resulting expressions turned out to be very simple and easy to deal with.
\end{abstract}


\maketitle

\section{Introduction}
Since Dirac's work on the relativistic equation of quantum mechanics, the interest in the Clifford algebra has become stronger and stronger due to the fact that the Dirac matrices satisfy this algebra. There is even a model based upon the self-interaction of these matrices that gives an explanation of why the neutrino has no electric charge \cite{Novello:1973jd}. The generators of this algebra can be constructed through products of the Dirac matrices and every so often we come across products between two different generators. An example of such a situation is the Dirac equation in a curved spacetime, where one usually faces products like $\gamma^A\gamma^{[B}\gamma^{C]}$, with $\gamma^A$ being the Dirac matrices.  There is no doubt that the more identities concerning those generators we know, the easier our calculation becomes. However, as far as I know no explicit expression for these products has been given in the literature. In order to fill this gap, I present here a full list with all possibles products between two different generators. I will not show the calculation that led to the identities, nevertheless, I shall convince the reader of their validity by writing down part of an algorithm developed in Maple 14 that I used to check all the identities presented here.

This article is organized as follows. Sec. \ref{s1392011a} is devoted to the notation and conventions adopted here. In Sec. \ref{a1392011b} the products of the generators are given in terms of the generators themselves, while Sec. \ref{a1392011c} is dedicated to the algorithm mentioned before. Some final comments are left to Sec. \ref{a1392011d}.

\section{Notation and conventions} \label{s1392011a}
Throughout this paper capital Latin letters will represent tetrad indices, which is the only type of index that will be used here, and all the results will be written in the tetrad basis. In this basis, the components of the metric will be denoted by $\eta_{AB}=\eta^{AB}=diag (1,-1,-1,-1)$.

Following the standard notation, I use ``$[|\ldots |]$'' to indicate the antisymmetric part of a tensor. For instance,  $\gamma^{[A|}\gamma^B\gamma^{|C]}=\frac{1}{2}\left(\gamma^A \gamma^B\gamma^C-\gamma^C \gamma^B\gamma^A  \right)$. When no vertical bar is present, one must antisymmetrize all indices inside the brackets. For example, $\gamma^{[A}\gamma^B\gamma^{C]}=\frac{1}{6}(\gamma^A \gamma^B\gamma^C+\gamma^C \gamma^A\gamma^B+\gamma^B \gamma^C\gamma^A-\gamma^A \gamma^C\gamma^B-\gamma^B \gamma^A\gamma^C-\gamma^C \gamma^B\gamma^A)$.

The Levi-Civita alternating symbol will be denoted by $\epsilon_{ABCD}$, where $\epsilon_{0123}=\epsilon^{0123}=+1$. Notice that this is just a symbol, not a component of a tensor. Besides, $\epsilon^{ABCD}\neq \eta^{AL}\eta^{BM}\eta^{CN}\eta^{DO}\epsilon_{LMNO}$. Nonetheless, we can define a pseudo-tensor through the identification $\varepsilon_{ABCD}\equiv \epsilon_{ABCD}$. In this case, we have $\varepsilon^{ABCD}= \eta^{AL}\eta^{BM}\eta^{CN}\eta^{DO}\varepsilon_{LMNO}=\eta^{-1}\epsilon^{ABCD}=-\epsilon^{ABCD}$, where $\eta$ is the determinant of the metric.  

There are many ways to represent the generators of the Clifford algebra. Nevertheless, I will stick to $\{\mathbb{I},\gamma^A,\gamma^{[A}\gamma^{B]}, \gamma^{[A}\gamma^B\gamma^{C]}, \gamma^{(5)}\}$, where $\gamma^{(5)}=\gamma^{(0)}\gamma^{(1)}\gamma^{(2)}\gamma^{(3)}$; the parenthesis is to emphasize that the ``gammas'' are written in the tetrad basis, which is assumed to be the standard Dirac matrices in four dimensions. These matrices satisfy $\gamma^A\gamma^B+\gamma^B\gamma^A=2\eta^{AB}\mathbb{I}$, where the unit matrix $\mathbb{I}$ will sometimes be omitted.

\section{Product of the generators of the Clifford algebra} \label{a1392011b}
The list below shows all possible combinations of the product between two generators of the Clifford algebra.
\begin{widetext}
\begin{eqnarray}
\gamma^A \gamma^B=\gamma^{[A} \gamma^{B]}+\eta^{AB}, \label{1292011a}
\\
\gamma^E\gamma^{[A} \gamma^{B]}=\gamma^{[E} \gamma^A\gamma^{B]}+\eta^{EA}\gamma^B-\eta^{EB}\gamma^A, \label{1292011b}
\\
\gamma^{[A} \gamma^{B]}\gamma^E=\gamma^{[E} \gamma^A\gamma^{B]}-\eta^{EA}\gamma^B+\eta^{EB}\gamma^A, \label{1292011c}
\\
\gamma^E \gamma^{[A} \gamma^B\gamma^{C]}= -\varepsilon^{EABC}\gamma^{(5)}        +\eta^{EA}\gamma^{[B}\gamma^{C]}+\eta^{EB}\gamma^{[C}\gamma^{A]}+\eta^{EC}\gamma^{[A}\gamma^{B]}, \label{1292011d}
\\
\gamma^{[A} \gamma^B\gamma^{C]}\gamma^E =\varepsilon^{EABC}\gamma^{(5)}+\eta^{EA}\gamma^{[B}\gamma^{C]}+\eta^{EB}\gamma^{[C}\gamma^{A]}+\eta^{EC}\gamma^{[A}\gamma^{B]}, \label{1292011e}
\\
\gamma^E\gamma^{(5)}=-\gamma^{(5)}\gamma^E=\frac{1}{3!}\varepsilon^E_{\ \ ABC}\gamma^{[A}\gamma^B\gamma^{C]}, \label{1292011e}
\\
\gamma^{[A}\gamma^{B]} \gamma^{[D}\gamma^{E]}=-\varepsilon^{DEAB}\gamma^{(5)}+\varepsilon^{AB\quad \,  H}_{\ \ \ \ [F|}\varepsilon^{DE}_{\ \ \ \ |G]H}\gamma^{[F}\gamma^{G]}+\eta^{BD}\eta^{AE}-\eta^{DA}\eta^{BE}, \label{1292011f}
\\
\gamma^{[D}\gamma^{E]} \gamma^{[A}\gamma^B\gamma^{C]}=\frac{1}{3}\varepsilon^{[D|ABC}\varepsilon^{|E]}_{\ \ FGH}\gamma^{[F}\gamma^G\gamma^{H]}+\varepsilon^{ABCF}\varepsilon^{DE}_{\quad \ HF}\gamma^H, \label{1292011g}
\\
\gamma^{[A}\gamma^B\gamma^{C]} \gamma^{[D}\gamma^{E]} =-\frac{1}{3}\varepsilon^{[D|ABC}\varepsilon^{|E]}_{\ \ FGH}\gamma^{[F}\gamma^G\gamma^{H]}+\varepsilon^{ABCF}\varepsilon^{DE}_{\quad \ HF}\gamma^H, \label{1292011h}
\\
\gamma^{[D}\gamma^{E]}\gamma^{(5)}=\gamma^{(5)}\gamma^{[D}\gamma^{E]}=  \frac{1}{2}\varepsilon^{ED}_{\quad \ AB}\gamma^{[A}\gamma^{B]}, \label{1292011i}
\\
\gamma^{[H}\gamma^F\gamma^{G]}\gamma^{[A}\gamma^B\gamma^{C]}=\varepsilon^{ABC}_{\quad \quad [D|} \varepsilon^{HFG}_{\quad \quad |E]} \gamma^{[E}\gamma^{D]}+\varepsilon^{HFGD}\varepsilon^{ABC}_{\quad \quad D}, \label{1292011j}
\\
\gamma^{[H}\gamma^F\gamma^{G]}\gamma^{(5)}=-\gamma^{(5)}\gamma^{[H}\gamma^F\gamma^{G]}=\varepsilon_A^{\ \ HFG}\gamma^A, \label{1292011l}
\\
\gamma^{(5)}\gamma^{(5)}=-\mathbb{I}. \label{1292011m}
\end{eqnarray}
\end{widetext}
Some additional identities that the reader may verify easily and that can be used with the previous ones are:

\begin{widetext}
\begin{eqnarray}
\varepsilon^{AB\quad H}_{\quad \ [F|} \varepsilon^{DE}_{\quad \ |G]H}\gamma^{[F}\gamma^{G]}=\eta^{EA}\gamma^{[B}\gamma^{D]}+\eta^{EB}\gamma^{[D}\gamma^{A]}+\eta^{DA}\gamma^{[E}\gamma^{B]}+\eta^{DB}\gamma^{[A}\gamma^{E]}
\\
\frac{1}{3}\varepsilon^{[D|ABC}\varepsilon^{|E]}_{\ \ FGH}\gamma^{[F}\gamma^G\gamma^{H]}=\eta^{EA}\gamma^{[D}\gamma^B\gamma^{C]}+\eta^{DA}\gamma^{[E}\gamma^C\gamma^{B]}+\eta^{EC}\gamma^{[D}\gamma^A\gamma^{B]}
\nonumber \\
+\eta^{DC}\gamma^{[A}\gamma^E\gamma^{B]}+\eta^{DB}\gamma^{[E}\gamma^A\gamma^{C]}+\eta^{EB}\gamma^{[D}\gamma^C\gamma^{A]}
\\
\varepsilon^{ABCF}\varepsilon^{DE}_{\quad \ HF}\gamma^H=(\eta^{DB}\eta^{EA}-\eta^{DA}\eta^{EB})\gamma^C+(\eta^{DA}\eta^{EC}-\eta^{DC}\eta^{EA})\gamma^B
\nonumber \\
+(\eta^{DC}\eta^{EB}-\eta^{DB}\eta^{EC})\gamma^A
\\
\varepsilon^{ABC}_{\quad \quad [D|} \varepsilon^{HFG}_{\quad \quad |E]} \gamma^{[E}\gamma^{D]}=(\eta^{HC}\eta^{BF}-\eta^{CF}\eta^{HB})\gamma^{[G}\gamma^{A]}+(\eta^{HC}\eta^{BG}-\eta^{CG}\eta^{HB})\gamma^{[A}\gamma^{F]}
\nonumber \\
+(\eta^{CG}\eta^{BF}-\eta^{CF}\eta^{BG})\gamma^{[A}\gamma^{H]}+(\eta^{AG}\eta^{HB}-\eta^{HA}\eta^{BG})\gamma^{[C}\gamma^{F]}
\nonumber \\
+(\eta^{AF}\eta^{HB}-\eta^{HA}\eta^{BF})\gamma^{[G}\gamma^{C]}+(\eta^{AF}\eta^{BG}-\eta^{AG}\eta^{BF})\gamma^{[C}\gamma^{H]}
\nonumber \\
+(\eta^{CG}\eta^{HA}-\eta^{HC}\eta^{AG})\gamma^{[B}\gamma^{F]}+(\eta^{CF}\eta^{HA}-\eta^{HC}\eta^{AF})\gamma^{[G}\gamma^{B]}
\nonumber \\
+(\eta^{CF}\eta^{AG}-\eta^{CG}\eta^{AF})\gamma^{[B}\gamma^{H]}
\\
\varepsilon^{HFGD}\varepsilon^{ABC}_{\quad \quad D}=\eta^{AH}(\eta^{BG}\eta^{CF}-\eta^{BF}\eta^{CG})+\eta^{AG}(\eta^{BF}\eta^{CH}-\eta^{BH}\eta^{CF})
\nonumber \\
+\eta^{AF}(\eta^{BH}\eta^{CG}-\eta^{BG}\eta^{CH}) \label{1592011a}
\\
\gamma^{[E}\gamma^A\gamma^B\gamma^{C]}=-\varepsilon^{EABC}\gamma^{(5)}
\end{eqnarray}
\end{widetext} 
To verify these identities, one may use \footnote{For more detail, see Ref. \cite{Inverno}, p. 92.}
\begin{equation}
\epsilon^{ABCD}\epsilon_{EFGH}=det\left(\begin{array}{cccc}
 \delta^A_E&\delta^B_E &\delta^C_E &\delta^D_E \\
 \delta^A_F&\delta^B_F &\delta^C_F &\delta^D_F \\
 \delta^A_G&\delta^B_G &\delta^C_G &\delta^D_G \\
 \delta^A_H&\delta^B_H &\delta^C_H &\delta^D_H
\end{array}\right).
\end{equation}

\section{Checking the previous identities} \label{a1392011c}
Instead of performing the calculations which lead to the identities shown in the previous section, I give here an algorithm developed in Maple 14 that was used to verify the validity of the identities (\ref{1292011a})-(\ref{1592011a}). This algorithm may look like naive, but it is sufficient for what we need. Its principal is simple: it takes the left-hand side of the identity and subtract it by its right-hand one. If the calculations are right, then the result is a $4\times 4$ null matrix --- which is not shown. On the other hand if there is something wrong, the computer shows ``it is not right''' and indicates the components that failed. Since it is not worth writing the whole algorithm, I will write down only the part that I used to check (\ref{1292011j}). 

Opening the package ``physics''\\
\verb|> with(Physics):|\\ 
Choosing a representation for the Dirac matrices\footnote{Be aware that the $\gamma^{(0)}$ version of the package ``Physics'' ( Male 14) for Majorana representation is wrong.}\\
\verb|> Setup(Dgammarepresentation = standard):|\\
Defining the contravariant Dirac matrices\\
\verb|> DiracMatrix:=A-> convert(Dgamma[A],Matrix):|\\
Defining the covariant version\\
\verb|> DiracMatrix_down:=A-> piecewise(A=0,|\\
\verb| convert(Dgamma[A],Matrix),|\\
\verb|-convert(Dgamma[A],Matrix)):|\\
Defining a list with sixteen zeros\\
\verb|>nula:=[0,0,0,0,0,0,0,0,0,0,0,0,0,0,0,0];|\\
Defining the metric tensor (Note that, in the tetrad basis, we need not make any distinction between covariant and contravariant metric).\\
\verb|>eta:=(A,B)->piecewise(A=0 and B=0,1, A=B and|\\
\verb| A<>0,-1,0);|\\
Defining the Levi-Civita alternating symbols (Notice that these symbols are not tensor, unlike the command ``LeviCivita[A,B,C,d]'' defined in Maple, which is a pseudo-tensor; note also that ``epsilon[0,1,2,3]=LeviCivita[1,2,3,0]=1'').\\
\verb|> Setup(signature = `+`);|\\
\verb|> for A from 0 to 3 do  for B from 0 to 3 do|\\
\verb|  for C from 0 to 3 do for d from 0 to 3 do|\\
\verb|> epsilon[d,A,B,C]:=LeviCivita[A,B,C,d]:|\\
\verb|> end do end do end do end do:|\\
Coming back to spacetime signature\\
\verb|> Setup(signature = `-`);|\\
Defining the contravariant version of the ``two-gamma generator''\\
\verb|> two:=(H,F)->1/2*(DiracMatrix(H).DiracMatrix(F)|\\
\verb|-DiracMatrix(F).DiracMatrix(H)):|\\
Defining the covariant version of the ``two-gamma generator''\\
\verb|> two_down:=(H,F)-> piecewise(H=0 or F=0,|\\
\verb| -two(H,F),two(H,F)):|\\
Defining the contravariant version of the ``three-gamma generator''\\
\verb|> three:=(H,F,G)->|\\
\verb| 1/6*(DiracMatrix(H).DiracMatrix(F).DiracMatrix(G)|\\
\verb|+DiracMatrix(G).DiracMatrix(H).DiracMatrix(F)|\\
\verb|+DiracMatrix(F).DiracMatrix(G).DiracMatrix(H)|\\
\verb| -DiracMatrix(H).DiracMatrix(G).DiracMatrix(F)|\\
\verb|-DiracMatrix(F).DiracMatrix(H).DiracMatrix(G)|\\
\verb|-DiracMatrix(G).DiracMatrix(F).DiracMatrix(H)  ):|\\
Defining the covariant version of the ``three-gamma generator''\\
\verb|> three_down:=(H,F,G)-> piecewise(H=0 or F=0 or|\\
\verb| G=0,three(H,F,G),-three(H,F,G)):|\\
Defining the contravariant version of the ``four-gamma generator''\\
\verb|> gamma5:=DiracMatrix(0).DiracMatrix(1)|\\
\verb|.DiracMatrix(2).DiracMatrix(3):|\\
Defining a second contravariant version of the ``four-gamma generator''\\
\verb|> four:=(H,F,G,E)->epsilon[H,F,G,E]*gamma5:|\\
Checking the identity (\ref{1292011j})\\
\verb|> for H from 0 to 3 do for F from 0 to 3 do|\\
\verb| for G from 0 to 3 do for A from 0 to 3 do|\\
\verb| for B from 0 to 3 do for C from 0 to 3 do|\\
\verb|> zero:=(H,F,G,A,B,C)-> three(H,F,G).three(A,B,C)|\\
\verb|-( 1/2*add(add( (epsilon[A,B,C,d]*epsilon[H,F,G,E]|\\
\verb|-epsilon[A,B,C,E]*epsilon[H,F,G,d] )*two_down(E,d),|\\
\verb| E=0..3),d=0..3)|\\
\verb|+add(add(eta(d,L)*epsilon[H,F,G,d]*epsilon[A,B,C,L],|\\
\verb|d=0..3),L=0..3)   );|\\
\verb|> s:=convert(zero(H,F,G,A,B,C),list):|\\
\verb|> if s<>nula then print("it is not right",|\\
\verb| H,F,G,A,B,C); print(zero(H,F,G,A,B,C)); end if;|\\
\verb|> end do end do end do end do end do end do;|\\

This algorithm is supposed to show nothing if the identity is right. The only difference between this algorithm and the ones for the other identities is the calculation right after ``checking the identity (\ref{1292011j})''.

\section{Final remarks} \label{a1392011d}
A natural continuation of this work is to find the products of the generators in a $n$-dimensional spacetime. However, this does not seem to be an easy task because the number of products would not be defined in this case.

\end{document}